\documentclass[prl,twocolumn,aps,amssymb,showpacs,superscriptaddress]{revtex4}
\usepackage{graphicx}

\begin{document}

\title{Observation of an exceptional point in a chaotic optical microcavity}

\author{Sang-Bum Lee}
\affiliation{School of Physics and Astronomy, Seoul National University, Seoul 151-747, Korea}
\affiliation{Korea Research Institute of Standard and Science, Daejeon 305-600, Korea}
\author{Juhee Yang}
\author{Songky Moon}
\author{Soo-Young Lee}
\affiliation{School of Physics and Astronomy, Seoul National University, Seoul 151-747, Korea}
\author{Jeong-Bo Shim}
\affiliation{Max Planck Institute for the Physics of Complex Systems, N\"{o}thnitzer Str. 38, Dresden, Germany}
\author{Sang Wook Kim}
\affiliation{Department of Physics Education, Pusan National University, Busan 609-735, Korea}
\author{Jai-Hyung Lee}
\author{Kyungwon An}
\email{kwan@phya.snu.ac.kr}
\affiliation{School of Physics and Astronomy, Seoul National University, Seoul 151-747, Korea}

\date{\today}

\begin{abstract}
We present spectroscopic observation of an exceptional point or the transition point between diabatic crossing and avoided crossing of neighboring quasi-eigenmodes in a chaotic optical microcavity with a large size parameter. The transition to the avoided crossing was impeded until the degree of deformation exceeded a threshold deformation owing to the system's openness also enhanced by the shape deformation. As a result, a singular topology was observed around the exceptional point on the eigenfrequency surfaces, resulting in fundamental inconsistency in mode labeling.
\end{abstract}


\maketitle

Dielectric optical microcavities are widely used in various optoelectronics applications such as add/drop filters, low-threshold microlasers \cite{Spillane}, single-molecule sensors \cite{Armani2007}, tunable optical frequency comb \cite{comb} and optomechanical oscillators \cite{optmech} owing to the high quality factor $Q$ of supported modes. In addition, asymmetric optical microcavities have drawn much attention because they exhibit directional output emission of high $Q$ modes. They can also serve as a useful platform for investigating the correspondence between quasi-eigenstates and associated chaotic classical dynamics in mesoscopic systems due to the well-known one-to-one correspondence between the Schr\"{o}dinger equation and the Maxwell equation in billiard problems \cite{Noeckel97, Rex02, Gmachl98, Lee02, Gmachl02, Lee05, Lee07PRA, Tanaka07, shim08PRL}. Level dynamics of interacting modes have also been studied for tailoring output directionality and quality factors of associated modes \cite{Wiersig06PRA, Wiersig06PRL} in microcavities.
 
One of the key issues related to the level dynamics is an exceptional point (EP), a singular point generally existing in a parameter space of a non-Hermitian system \cite{Heiss00, Berry98, JPA09}. EPs have been observed in various systems such as acoustic systems \cite{Shuvalov00}, atoms in optical lattices \cite{Oberthaler96}, and complex atoms in laser fields \cite{Latinne95}. In coupled microwave cavities the EP was studied in terms of resonance mode distributions \cite{Dembowsk01, Dembowsk03}. Existence of an EP has been predicted for mesoscopic systems such as a Rydberg atom in a strong magnetic field \cite{hydrogen} and in a stadium-shape microcavity \cite{SYL08}, where classical chaos is also important. Yet their experimental verifications in these classically chaotic systems have not been reported.

In a recent work of Ref.\ \cite{Lee-PRA09}, we observed both diabatic crossing (DC) and avoided crossing (AC) in a chaotic optical microcavity (COM), depending on the degree of cavity-shape asymmetry. However, it was not possible to observe an EP, where DC and AC coalesce in a two-dimensional parameter space.  The problem was a finite spectral resolution, which prevented us from telling whether two modes were crossing each other or they were avoiding each other with a gap smaller than the spectral resolution. Similar problems were also found with indistinguishable static envelops of diabatic and avoided crossing modes in a high-$Q$ toroidal microcavity \cite{Carmon-PRL07}.

In this Letter we report the first observation of an EP in a high-Q asymmetric microcavity or COM with its boundary shape continuously variable. Our experiment done in a single COM, not in coupled cavities as before \cite{Dembowsk01, Dembowsk03}, was made possible by introducing an internal parameter, a quasi-continuous variable in a semiclassical regime of large size parameter, and is based on the fact that  two modes undergoing an AC exhibit fundamentally different output coupling signatures from those of DC modes. Moreover, we elucidate the resulting singular topology in the eigenenergy surfaces around the EP as an inherent source of impossibility of consistent mode labeling in open mesoscopic systems. 

Our experiment was performed with a COM or a two-dimensional microcavity formed by a liquid jet of ethanol (refractive index $m$=1.361 at 610 nm) dilutely doped with Rhodamine dye molecules. The boundary profile of the microcavity is approximated by $r(\phi)\simeq a(1+\eta\cos2\phi+\epsilon\eta^2\cos4\phi)$ in the polar coordinates with $a\simeq$14.9$\pm0.1\mu$m and $\epsilon=0.42\pm0.05$ \cite{Yang-RSI06, Moon-OE08}. The deformation parameter $\eta$ can be continuously varied from 0\% to 26\% at will. The size parameter, defined as $2\pi ma/\lambda$ with $\lambda$ the wavelength, is about 190 for $\lambda$=660 nm.

The COM was pumped by a cw argon-ion laser and the cavity-modified fluorescence (CMF) or lasing light from the COM was measured with a spectrometer. The polarization direction for both pumping and detection was parallel to the COM column \cite{Lee02, Lee07PRA, Lee-PRA09}.
There were observed five different mode sequences recurring with an interval or a free spectral range (FSR) of about 2.3 THz. Mode label $l$ is assigned to each mode sequence by the scheme explained in Ref.\ \cite{Lee-PRA09} and the recurring modes in a mode sequence are indexed by mode number $n$. 

\begin{figure}
\includegraphics[width=3.3in]{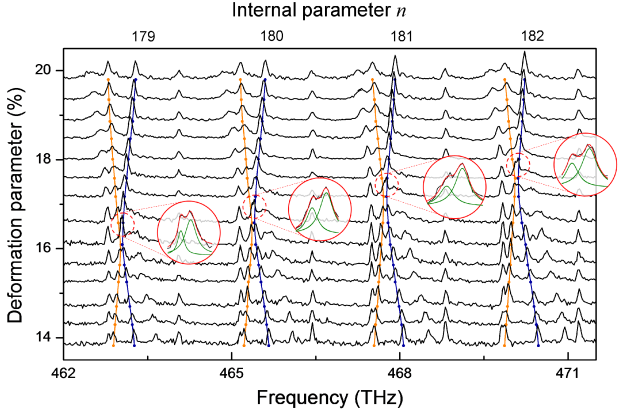}
\caption{Eigenfrequencies of $l$=2 (in orange) and $l$=4 (in blue) modes as a function of the deformation $\eta$ and the internal parameter $n$. The magnitude of AC increases as $\eta$ increases. The magnified view shows the high resolution spectra taken with a spectrometer of 0.05 nm resolution. Green curves are Lorentzian fits.}
\label{spectrum}
\end{figure}

In Fig.\ \ref{spectrum}, we observe $l$=2 and $l$=4 modes undergoing ACs as a function of $\eta$ at 
$n$=179, 180, 181 and 182, respectively, indicating that the separation of two modes can be adjusted continuously by changing $\eta$ for a fixed $n$. We also observe ACs as we move horizontally in Fig.\ \ref{spectrum} for a fixed $\eta$, for example $\eta$=0.176, from one FSR to another. This is equivalent to scanning the mode number $n$, a discrete internal parameter. The relative frequency between $l$=2 and 4 modes is changed by 0.083 THz  or 3.6\% (when they are well apart) of their FSRs when $n$ is shifted by one. From the locations of ACs,  ($n$, $\eta$)=(179, 0.162), (180, 0.167), (181, 0.172), (182, 0.176), we also find that the same 0.083 THz induced by $n$ shift can be continuously scanned by changing $\eta$ by about 0.005. Therefore, by coarse scanning of $n$ shift for a fixed $\eta$, followed by a fine adjustment of $\eta$ within 0.005, we can bring any two modes together, far beyond our spectrometer resolution ($\sim$50 GHz). Exact definition of the internal parameter $n$ is given below when we discuss Fig.\ \ref{topology}.

Before we examine the interaction between modes $l$=2 and 4 closely, let us recapitulate a theoretical background on mode-mode interaction. It was shown in Ref.\ \cite{Lee-PRA09} that the mode-mode interaction in a COM can be described by a 2-by-2 non-Hermitian symmetric Hamiltonian. Its diagonal elements, given by $E_j(n,\eta)=\nu_j(n,\eta)-i\gamma_j(\eta)$ ($j$=$a, b$) (with the Planck constant $h=1$), are the eigenvalues of quasi-eigenmodes when their eigenvalues are well separated, {\em i.e.}, effectively uncoupled. The imaginary part corresponds to the decay rate of the mode. The symmetric off-diagonal element, denoted by $C(\eta)$, is the mode-mode coupling constant induced by the nonintegrability or the cavity shape asymmetry while its dependence on the internal parameter $n$ is negligible in the small spectral range of interest. 

The uncoupled quasi-eigenmodes of different radial mode order have different FSRs. Thus, by shifting the internal parameter $n$ for a fixed $\eta$ followed by a fine adjustment of $\eta$ as explained above, we can bring any two uncoupled modes together and make the mode-mode coupling come into play. In this case the system has new eigenvalues $E_\pm$ given by $E_\pm=(E_a+E_b)/2\pm\sqrt{(E_a-E_b)^2/4+{C}^2}$. When $\nu_a=\nu_b(=\nu_0)$, the new eigenvalues are given by $E_\pm=\nu_0\pm (C^2-\gamma_-^2)^{1/2}-i\gamma_+$ with $\gamma_\pm=|\gamma_a\pm\gamma_b|/2$. When $C<\gamma_-$, a splitting occurs in the imaginary part of energy, corresponding to a crossing in the real part as we vary $n$. When $C>\gamma_-$, an AC, a splitting in the real part, would occur as $n$ is varied. This splitting is analogous to the normal mode splitting in coupled cavities \cite{Dembowsk01} or in atom-cavity systems \cite{cavity-qed}. However, in our COM the internal coupling $C$ is induced by the cavity shape asymmetry, related to chaotic ray transport between phase-space regions associated with the involved quasi-eigenmodes \cite{Lee-PRA09}.

\begin{figure}
\includegraphics[width=3.3in]{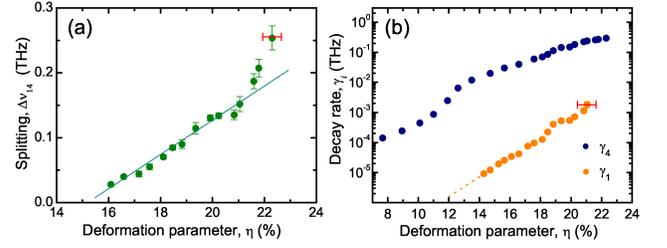}
\caption{(a) Magnitude $\Delta \nu_{14}$ of AC between two uncoupled states, $l_0$=1 and $l_0$=4 modes, as a function of $\eta$. Solid line is a linear fit. (b) Decay rates $\gamma_1$ and $\gamma_4$ of $l_0$=1,4 uncoupled states in the spectral region where AC or DC occurs. Dotted line for $\gamma_1$ is an extrapolation beyond spectral visibility.} 
\label{avoided-vs-deform}
\end{figure}

It should be noted that $l$=2 and $l$=4 modes in Fig.\ \ref{spectrum} correspond to the new quasi-eigenvalues $E_\pm$, the result of AC between uncoupled $j$=$a, b$ quasi-eigenmodes. We have systematically measured the sizes of AC between $l$=2 and $l$=4 modes for various cavity deformation ranging from 0.14 to 0.23. We can also obtain the decay rates or the half linewidths of the uncoupled states in the spectral region where AC or DC occurs. In fact, we have identified from the spectrum evolution that these uncoupled states have evolved from whispering gallery modes (WGMs) with radial mode order $l_0$=1 and 4 in a circular cavity as $\eta$ is gradually increased \cite{Lee-PRA09}. Thus we can label these uncoupled states $j$=$a, b$ above by the same $l_0$=1 and 4 as those of the original WGMs. The measured magnitude $\Delta \nu_{14}$ of AC between $l_0$=1 and 4 states and their decay rates $\gamma_1$ and $\gamma_4$ are summarized in Figs.\ \ref{avoided-vs-deform}(a) and \ref{avoided-vs-deform}(b), respectively.

It is interesting to note that the observed $\Delta\nu_{14}$'s appear to be well fit by a straight line with an $\eta$-axis offset, indicating there exist a threshold deformation for the splitting. 
The transition from DC to AC is suppressed up to a threshold deformation due to the openness which is also enhanced by system's nonintegrability. This feature is quite interesting since the nonintegrability in an open chaotic billiard induces both coupling and openness (the latter summed up by decay rates) to grow and it is not obvious how rapidly the coupling and the decay rates would increase as the degree of nonintegrability $\eta$ rises. In our example of $l_0$=1 and 4, the coupling (initially zero) catches up the decay rates (initially finite) as $\eta$ grows. Contrarily, in a closed system an AC would occur as long as $\eta$$>$$0$ however small it is since there is no decay by definition. 

From the $\eta$-axis offset of a linear fit in Fig.\ \ref{avoided-vs-deform}(a), one may expect that the two modes $l_0$=1 and 4 would undergo a transition from AC to DC, or vice versa, at $\eta\sim$ 0.15. However, the linewidth ($<$3 GHz) of these modes become much narrower than our spectral resolution once $\eta$ is decreased below 0.16, and thus we cannot determine the actual value of threshold deformation by simply noting the disappearance of the splitting between the two modes. In fact, this problem would persist no matter how high the spectral resolution is unless it is infinite.

One can solve this rather inherent problem by realizing the fact that two modes undergoing a DC maintain their original linewidths whereas two modes undergoing an AC share a common linewidth given by a geometric mean of the original linewidths. This fundamental difference is then translated to substantially different output coupling efficiencies for those two cases as to be seen below. Here the output coupling efficiency is a measure of the mode strength seen in the spectrum, given by the ratio $\epsilon_j$$=$$\gamma_j / (\gamma_j + \gamma_{\rm abs})$ for the $j$th mode with $\gamma_{\rm abs}$ the absorption rate of the cavity medium \cite{Chylek91}. 

\begin{figure}
\includegraphics[width=3.3in]{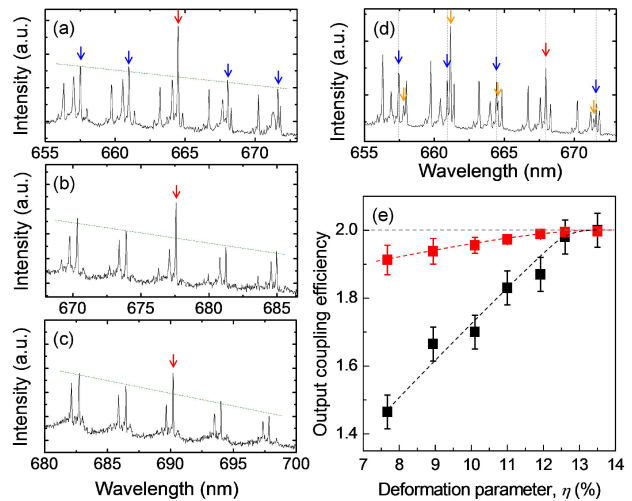}
\caption{ CMF spectra for the case of (a) $\eta$=0.125, (b) $\eta$=0.102 and (c) $\eta$=0.078. Orange and blue arrows indicate $l_0$=1, 4 modes, respectively. The output coupling efficiency of $l_0$=1 modes is too small for these modes to be observed in (a)-(c). Red arrows indicate where $l_0$=1 and 4 modes appear to be overlapped. Deformation $\eta$ was fine tuned within 0.005 to maximize the peak. (d) Pump power is increased above a lasing threshold for $l_0$=1 mode at 661 nm with $\eta$=0.120. $l_0$=1 modes are clearly seen. (e) Observed output coupling efficiency (black squares) and the expected output coupling efficiency $\epsilon_{AC}$ (red squares). Slight reduction of $\epsilon_{AC}$ itself is mostly due to the decrease of decay rates of the involved modes. Black dotted line is a spline fit for visual guidance.} 
\label{fig3}
\end{figure}

\begin{figure}
\includegraphics[width=3.3in]{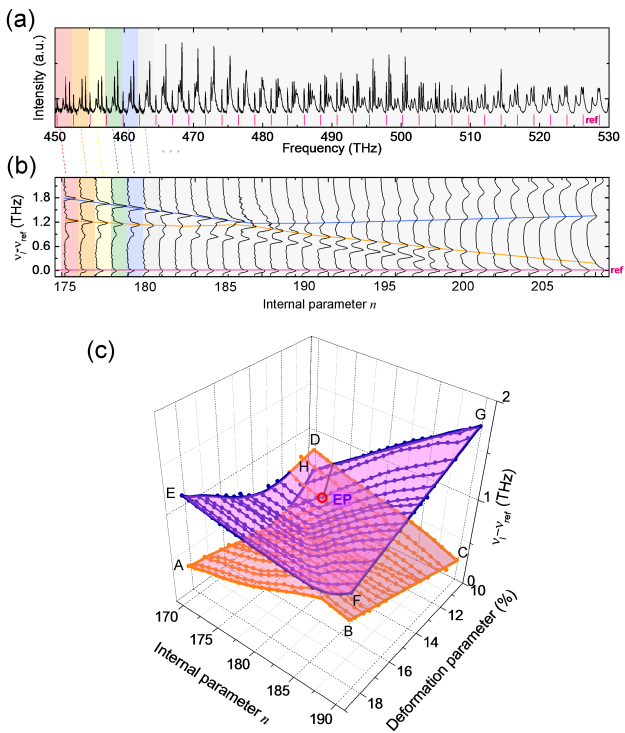}
\caption{(a) The observed spectrum for $\eta$=0.187. (b) The relative frequencies of observed quasi-eigenmodes with respect to the reference frequency of the same $n$ as a function of $n$. (c) Eigenfrequency surfaces of two coupled states $l$=2 (orange dots) and $l$=4 modes (blue dots) in $n$-$\eta$ parameter space show a complex-square-root-like topology with a branch-point singularity at $(n_0,\eta_0)$=(175, 0.125).}
\label{topology}
\end{figure}

For $l_0$=1 and 4 modes, we have $(\gamma_1, \gamma_4, \gamma_{\rm abs})\sim (2$$\times$$10^{-6}, 4$$\times$$10^{-3}, 2$$\times$$10^{-5})$ THz and thus $\gamma_1\ll\gamma_{abs}\ll\gamma_4$ in the spectral region where the transition is expected, and therefore $\epsilon_1\sim\gamma_1/\gamma_{abs}\ll$1 and $\epsilon_4\sim$1. This indicates that $l_0$=1 mode is hardly visible while $l_0$=4 mode has almost maximum visibility in the spectrum when these two modes are well separated. This behavior can be seen in Figs.\ \ref{fig3}(a)-\ref{fig3}(c), where $l_0$=1 modes are not visible in the fluorescence spectra whereas they become prominent when the pump power is increased beyond its lasing threshold as seen in Fig.\ \ref{fig3}(d). 

When an AC occurs $(C>\gamma_-\sim \gamma_4 /2)$, the two modes form a doublet with each having a decay rate of $\gamma_+ \sim \gamma_4 /2$. The output coupling efficiency of each is then given by $\epsilon_+=\gamma_+ / (\gamma_+ + \gamma_{abs}) \sim 1$. When the two modes barely undergo an AC, they appear to be overlapped into a single peak in the spectrum with its output coupling efficiency given by $\epsilon_{AC}=2\epsilon_+$. This effect is clearly observed in Fig.\ \ref{fig3}(a), where the peak at 664.5 nm is twice larger than the other peaks ($l_0$=4 modes) marked by blue arrows, indicating an AC takes place there between $l_0$=1 and 4 modes.

When the two modes undergo a crossing ($C \le \gamma_- $), on the other hand, decay rate of each mode is modified from its original value as $\gamma\prime_1=\gamma_+ - (\gamma_-^2-C^2)^{1/2}$ and $\gamma\prime_4=\gamma_+ + (\gamma_-^2-C^2)^{1/2}$. The resulting output coupling efficiency when the two modes are overlapped is then given by $\epsilon_{DC} =\gamma\prime_1/ (\gamma\prime_1 + \gamma_{abs})+ \gamma\prime_4 / (\gamma\prime_4 + \gamma_{abs}) \le 2\epsilon_+$, where the equality holds when $C=\gamma_-$ or at the transition point. Particularly, if $C\ll\gamma_-$ , $\gamma\prime_1 \sim \gamma_1$ and $\gamma\prime_4 \sim \gamma_4$, and the corresponding output coupling efficiency is approximately given by $\epsilon_{DC} \sim$1. Therefore, the transition from AC to DC is signaled by a substantial reduction of the output coupling efficiency from $2\epsilon_+$, as seen in Figs.\ \ref{fig3}(b) and \ref{fig3}(c), when two modes appear to be overlapped.

We examined the output coupling efficiency (black squares) for $l_0$=1 and 4 modes when they appear to be overlapped. We observed that this coupling efficiency starts to deviate noticeably from the expected output coupling efficiency $\epsilon_{AC}$ (red squares) as the deformation parameter is reduced below 0.125 as shown in Fig.\ \ref{fig3}(e). From this observation we conclude that the two modes undergo a transition from AC to DC, or vice versa, at $\eta_0$=0.125$\pm$0.005. 

A parameter-space point at which the transition from DC to AC takes place or $E_{\pm}$ collapse to one is an EP, a topological singular point. The singular nature of the EP is revealed when we examine the eigenfrequency surfaces of $l$=2 and 4 modes in a $n$-$\eta$ parameter space. The eigenfrequency surfaces $E_\pm(n,\eta)$ are constructed in the following way. We first define reference frequencies as the resonance frequencies of $l_0$=3 WGM in a circular cavity whose round trip length is the same as that of the COM under investigation, as shown in Fig.\ \ref{topology}(a). The spectrum is then evenly divided into segments with each 
being as wide as one free spectral range of the $l_0$=3 WGM so that each segment is indexed as $n$ of that WGM. This $n$ is precisely the internal parameter used throughout this work. Its value, representing the number of wavelengths fitting the orbit associated with the mode, is as large as the size parameter. The relative frequencies of observed quasi-eigenmodes assigned with $n$, with respect to the reference frequency of the same $n$, are then plotted as a function of $n$ as shown in Fig.\ \ref{topology}(b). By repeating this procedure for other $\eta$ and then combining the results as a function of $n$ and $\eta$, we finally obtain the eigenfrequency surfaces $E_\pm(n,\eta)$ as shown in Fig.\ \ref{topology}(c). The resulting surfaces exhibit a complex-square-root-function-like topology with a branch-point singularity at the EP located approximately at $(n_0,\eta_0)$=(175, 0.125), where both an AC and a DC between $l_0$=1 and 4 modes coalesce.

The singular topology of the eigenenergy surfaces around the EP result in a fundamental inconsistency in assigning mode labels to quasi-eigenmodes. In order to illustrate this point, let us consider a cyclic variation of ($n$, $\eta$) as shown in Fig.\ \ref{topology}(c): (170, 0.18)$\rightarrow$(190, 0.18)$\rightarrow$(190, 0.10)$\rightarrow$(170, 0.10)$\rightarrow$(170, 0.18) enclosing the EP. If we choose $l$=2 mode at (170, 0.18) and follow the mode under this cyclic variation, we end up with a different $l$=4 mode in the end after traversing A$\rightarrow$B$\rightarrow$C$\rightarrow$D$\rightarrow$E on the energy surface. What happens is that the mode label abruptly changes from $l$=2 to $l$=4 when we pass by the EP during adiabatic process of increasing $\eta$ (D$\rightarrow$E), and moreover, there is no way to avoid this inconsistency however differently we label the modes. We have to perform the cyclic variation once more, traversing E$\rightarrow$F$\rightarrow$G$\rightarrow$H$\rightarrow$A around the EP, in order to come back to the same starting mode. This consideration reveals a fundamental inconsistency in assigning mode labels to quasi-eigenmodes in nonintegrable open systems. This ambiguity is a direct consequence of the singular topology around the EP.

In conclusion, we have observed an EP or the transition point between DC and AC in a COM by utilizing different output coupling efficiencies in those two cases. The observed quasi-eigenfrequencies of interacting modes exhibit a branch-point topology, which is the very origin of impossibility of consistent mode labeling in these open chaotic systems. Our spectroscopic method using output coupling efficiency can be applied to other optical and optomechanical systems to resolve DC/AC static mode envelops \cite{Carmon-PRL07}.

This work was supported by NRL and WCU Grants. SWK was supported by KRF Grant (2008-314-C00144). SYL was supported by BK21 program.

\end{document}